\documentclass{aa}
\usepackage{float}
\usepackage{graphicx}
\usepackage{natbib}
\usepackage{txfonts}
\begin{document}

   \title{Kinematic data rebuild the Nuclear star cluster as the most metal rich region of the Galaxy}

  \author{F. Nogueras-Lara
          \inst{1}             
          }

   \institute{
    Max-Planck Institute for Astronomy, K\"onigstuhl 17, 69117 Heidelberg, Germany
              \email{nogueras@mpia.de}                                 
       }
   \date{}

 
  \abstract
   {The Galactic centre (GC) is located at only 8 kpc from Earth and constitutes a unique template to understand Galactic nuclei. Nevertheless, the high crowding and extinction towards the GC hamper the study of its main stellar components, the nuclear stellar disc (NSD) and the nuclear star cluster (NSC).}
   {Recent work has suggested that the NSD and the NSC can be distinguished along the line of sight towards the NSC via the different extinction of their stars. This motivated us to analyse the proper motion, radial velocity, and the metallicity distributions of the different extinction groups.} 
    {We use photometric, kinematic, and metallicity data to distinguish between probable NSD and NSC stars in a region centred on the NSC.}
   {We detected two different extinction groups of stars and obtained significantly different proper motion distributions for each of them, in agreement with the expected kinematics for the NSD and the NSC. We derived radial velocity maps that appear to be different for the NSD and the NSC. We also found different metallicities for each of the components, with the largest one measured for the most extinguished group of stars. We obtained that the metallicity distribution of each extinction group is best fitted by a bimodal distribution, indicating the presence of two metallicity components for each of them (a broad one slightly below solar metallicity, and a more metal rich narrower one, that is largest for the high extinction group of stars).}  
   {We conclude that both extinction groups are distinct GC components with different kinematics and metallicity, and correspond to the NSD and the NSC. Therefore, it is possible to distinguish them via their different extinction. The high mean metallicity, $[M/H]\sim0.3$\,dex, obtained for the NSC metal rich stars, supports that the NSC is arguabily the most metal rich region of the Galaxy.}


   \maketitle
%

\section{Introduction}

The Milky Way's centre is the closest galaxy nucleus and the only one where we can resolve individual stars down to milliparsec scales. Besides the supermassive black hole, Sagittarius A*, two main stellar structures outline the Galactic centre (GC): (1) the nuclear star cluster (NSC), a massive  stellar cluster \citep[$\sim2.5\times10^7$\,M$_\odot$, e.g.][]{Launhardt:2002nx,Schodel:2014bn,Feldmeier:2014kx} placed at the heart of the Galaxy with an effective radius of $\sim 5$\,pc \citep[e.g.][]{Graham:2009lh,Schodel:2011ab,Feldmeier-Krause:2017kq,gallego-cano2019}, and (2) the nuclear stellar disc (NSD), a much larger stellar structure \citep[with a radius $\sim200$\,pc, e.g.][]{Launhardt:2002nx,Nishiyama:2013uq,gallego-cano2019,Sormani:2020aa,Sormani:2022wv} surrounding the NSC, and partially overlapping with the dense gas from the central molecular zone \citep[e.g.][]{Henshaw:2022vl}.

In spite of being placed at only 8\,kpc from Earth \citep[e.g.][]{Gravity-Collaboration:2018aa,Do:2019aa}, the study of the GC is hampered by the high extinction and the extreme source crowding \citep[e.g.][]{Nishiyama:2008qa,Schodel:2010fk,Nogueras-Lara:2018aa,Nogueras-Lara:2020aa,Nogueras-Lara:2021wj}. Therefore, disentangling the GC components and analysing their main properties is a formidable challenge. Recent studies point towards a different stellar population and formation scenario for the NSC and the NSD \citep{Nogueras-Lara:2019ad,Schodel:2020aa,Nogueras-Lara:2021wm}. In this way, in spite of being dominated by old stars (more than 80\,\% of their stellar population is older than $8$\,Gyr), the NSC contains an intermediate age stellar population \citep[up to $15$\,\% of the stellar mass was formed $\sim 3$\,Gyr ago][]{Schodel:2020aa}, that is not present in the NSD. Analogously, a significant mass fraction of the NSD stars ($\sim5$\,\%) formed about $1$\,Gyr ago, corresponding to a time during which the NSC experienced no significant star formation activity  \citep{Pfuhl:2011uq,Nogueras-Lara:2019ad,Schodel:2020aa}. On the other hand, the metallicity of the NSC stars seems to be higher than that of the NSD \citep{Schultheis:2021wf,Feldmeier-Krause:2022vm}. 

Recently, \citet{Nogueras-Lara:2021wm} have shown that NSD and NSC stars are subject to significantly different reddening. Here, we follow up on their work by including kinematic and spectroscopic data. We find that the kinematics and metallicity distributions are different for each of the extinction groups, and that they are in agreement with the expected results for the NSD and the NSC.

\section{Data}

\subsection{Photometry}

         \begin{figure*}
   \includegraphics[width=\linewidth]{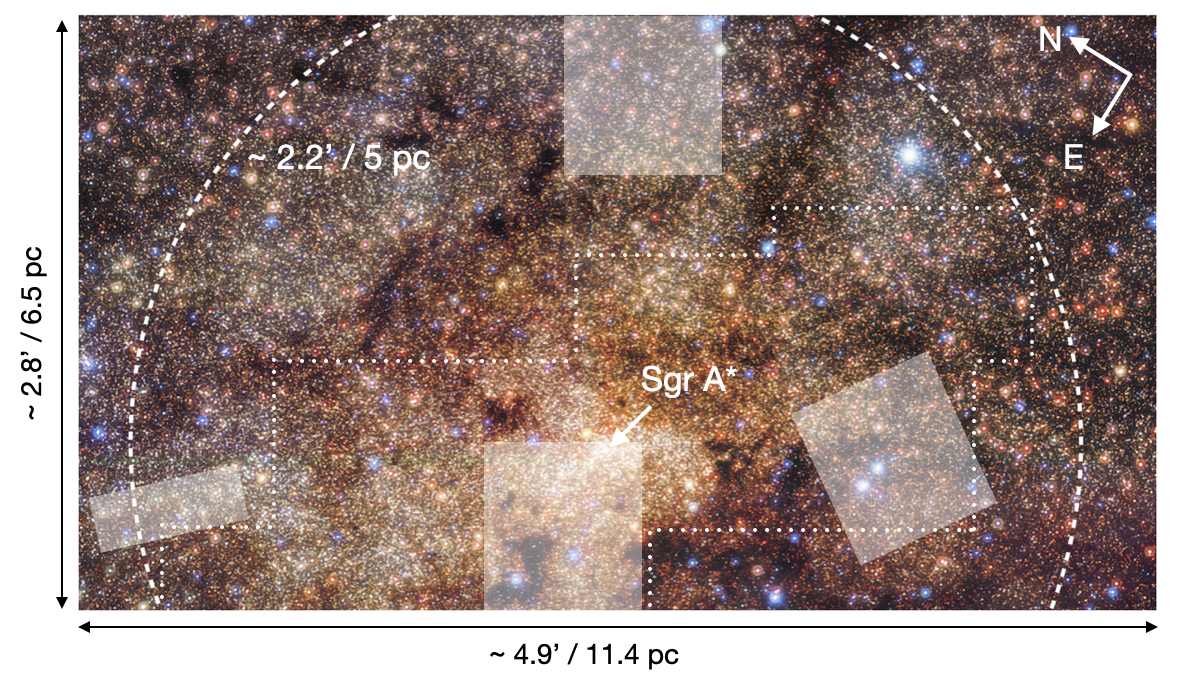}
   \caption{GALACTICNUCLEUS false-colour image of the analysed region using $JHK_s$ bands. The white dashed line shows the effective radius of the NSC. White shaded contours indicates regions where there is not coverage from the proper motion catalogue. The dotted contour shows the region with stars with known metallicities.}

   \label{GNS}
    \end{figure*}
    
The photometric observations used in this work were obtained with the HAWK-I instrument \citep[][]{Kissler-Patig:2008fr}, placed at the ESO Very Large Telescope in Chile (UT4). They are $J$ and $K_s$ data of a field of $\sim 2.8' \times 4.9'$ centred on the NSC (Fig.\,\ref{GNS}, coordinates 17$^h$ 45$^m$ 38$^s$, -29$^\circ$ 00$'$ 12$''$). The $J$ data belong to the GALACTICNUCLEUS survey \citep[GNS, ][]{Nogueras-Lara:2018aa,Nogueras-Lara:2019aa}, a high-angular resolution ($\sim 0.2''$) $JHK_s$ catalogue specially designed to observe the GC. The $K_s$-band data were obtained in 2013 and form part of a pilot study for the GNS survey \citep{Nogueras-Lara:2018aa} that was obtained under excellent observing conditions ($K_s$ seeing $\sim 0.4''$), allowing a photometry $\sim 1$\,mag deeper than for the corresponding GNS field (see Table in \citealt{Nogueras-Lara:2018aa}). 

To correct potential saturation problems affecting bright stars in $K_s$ \citep[e.g.][]{Nogueras-Lara:2019aa}, we used the SIRIUS/IRSF survey \citep{Nagayama:2003fk,Nishiyama:2006tx}  to replace the $K_s$ photometry of stars with $K_s< 11.5$\,mag \citep[for further details, see ][]{Nogueras-Lara:2019ad}.

\subsection{Proper motions}

We used a publicly available proper motion catalogue of the GC \citep{Shahzamanian:2019aa,Shahzamanian:2021wu}, that overlaps with the region analysed in this paper. This catalogue was specifically designed to study the GC and constitutes an unprecedented kinematic data set for the NSD region. The proper motions were computed using two data sets (the GNS $H$-band data, \citealt{Nogueras-Lara:2018aa,Nogueras-Lara:2019aa}, and the HST Paschen-$\alpha$ survey, \citealt{Wang:2010fk,Dong:2011ff}), with a timeline of $\sim 7-8$\,years between them. The high angular resolution of the used data sets \citep[$\sim0.2''$][]{Shahzamanian:2021wu}, allows the catalogue to supersede previous surveys covering the same area \citep[e.g. the VIRAC survey][]{Smith:2018aa}, that are limited to a narrow magnitude range in the analysed region, due to saturation and seeing-limited resolution \citep{Smith:2018aa, Shahzamanian:2021wu}.

Given the complex procedure of matching the data from the two different epochs, their different tiling patterns, and the different size of the detectors, the proper motion catalogue does not have a homogeneous coverage across its surveyed field \citep[see Fig.\,2 in ][]{Shahzamanian:2021wu}. Nevertheless, we checked that $\gtrsim 80\,\%$ of the region analysed in this study is properly covered as shown in Fig.\,\ref{GNS}. 

The computed proper motions are relative, not absolute, meaning that they were computed assuming an average zero motion between the stellar positions from the used reference catalogue to calculate the proper motions. This only means that there is an offset between proper motions considering an absolute relative frame and the ones in the catalogue. \citet{Shahzamanian:2021wu} carried out an in-detail comparison of their catalogue with absolute calibrated work \citep{Libralato:2021td}, and concluded that the proper motions perfectly agree once converted the reference frame. In any case, this paper does not pursue the calculation of absolute proper motions, but to analyse the proper motion distribution of the NSC and the NSD. Given that we are using the same pointings for the NSC and the NSD (instead of comparing different GC regions), the use of relative proper motions does not pose any problem.

\subsection{Metallicity and radial velocity}
\label{metal}

We used metallicity and radial velocity KMOS \citep{Sharples:2013aa} medium resolution spectra data for around 1,000 stars in the NSC region (see Fig.\,\ref{GNS}) obtained by \citet{Feldmeier-Krause:2017kq,Feldmeier-Krause:2020uv}. They derived the radial velocities using the IDL routine {\it pPXF} \citep{Cappellari:2004us}, using as templates the high-resolution spectra of \citet{Wallace:1996vy}. Then, they applied a full spectral fitting method with PHOENIX models \citep{Husser:2013uu} to derive the metallicities and other stellar parameters. The absolute metallicities were calibrated using existing empirical spectra for $[M/H] < 0.3$\,dex. This might produce an overestimation of metallicities $[M/H] > 0.3$\,dex. To check this, \citet{Schultheis:2021wf} derived the metallicities using the data from \citet{Feldmeier-Krause:2020uv} applying an alternative method using CO and NaI index, and empirical spectra calibrated for $[M/H] < 0.6$\,dex  \citep[for further details on the method see][]{Fritz:2020aa}. They concluded that there is not systematic bias for stars with $[M/H] < 0.5$\,dex, whereas stars with $[M/H] > 0.5$\,dex show significantly higher metallicities when applying the methodology in \citet{Feldmeier-Krause:2017kq}. On the other hand, this metallicity cut at $[M/H] = 0.5$\,dex also agrees with the most metal rich stars detected in the NSC when using high resolution spectroscopy \citep[e.g. ][]{Do:2015ve,Rich:2017rm,Thorsbro:2020uq}.

\section{Proper motion analysis}

\subsection{Disentangling the stellar populations}

\begin{figure}
   \includegraphics[width=\linewidth]{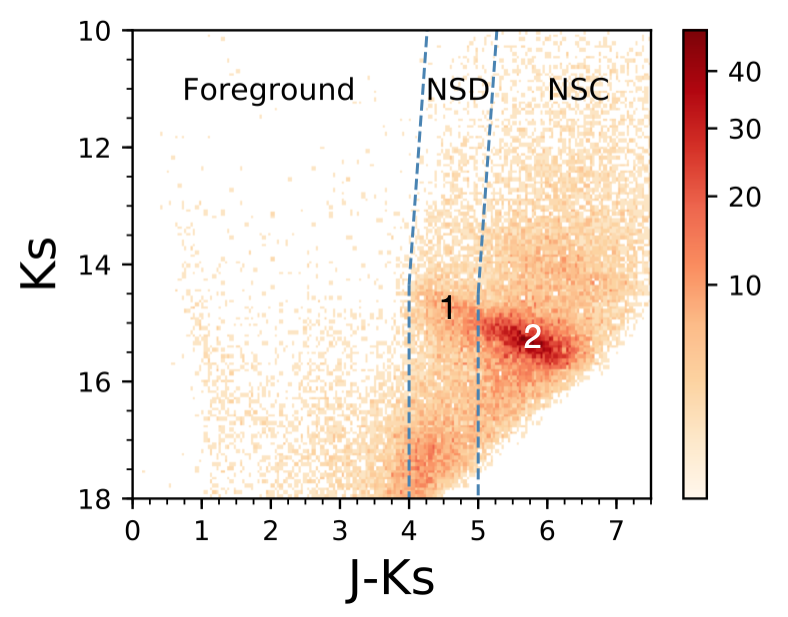}
   \caption{Colour-magnitude diagram $K_s$ versus $J-K_s$. The colour code corresponds to stellar densities using a power stretch scale, where the number of stars per bin is indicated with the colour bar. The blue dashed lines show the division between foreground, NSD, and NSC stars. The numbers 1 and 2 indicate the red clump bumps with different extinction that belong to the NSD, and the NSC, respectively.}

   \label{CMD}
    \end{figure}

Recent work by \citet{Nogueras-Lara:2021wm} used $HK_s$ photometry to show that the stellar populations from the NSC and the NSD can be identified along the line of sight towards the NSC via their significantly different extinction \citep[$A_{Ks\ NSD} \sim 1.7$\,mag, and $A_{Ks\ NSC} \sim 2.3$\,mag, ][]{Nogueras-Lara:2021wm}. In this work, we use $J$ data instead $H$ to improve the identification of the NSD and the NSC, given that the extinction is higher for shorter wavelengths \citep[e.g.][]{Nogueras-Lara:2021wj}, and thus the use of $J$-band data increases the difference between both stellar populations. Figure\,\ref{CMD} shows the colour-magnitude diagram (CMD) $K_s$ versus $J-K_s$, where we visually detect two red clump \citep[stars in their helium burning core sequence][]{Girardi:2016fk} bumps with different extinction, that correspond to the NSD (low extinction group) and the NSC (high extinction group), according to \citet{Nogueras-Lara:2021wm} (see also Fig.\,16 of \citealt{Nogueras-Lara:2018aa}).

\begin{figure}
   \includegraphics[width=\linewidth]{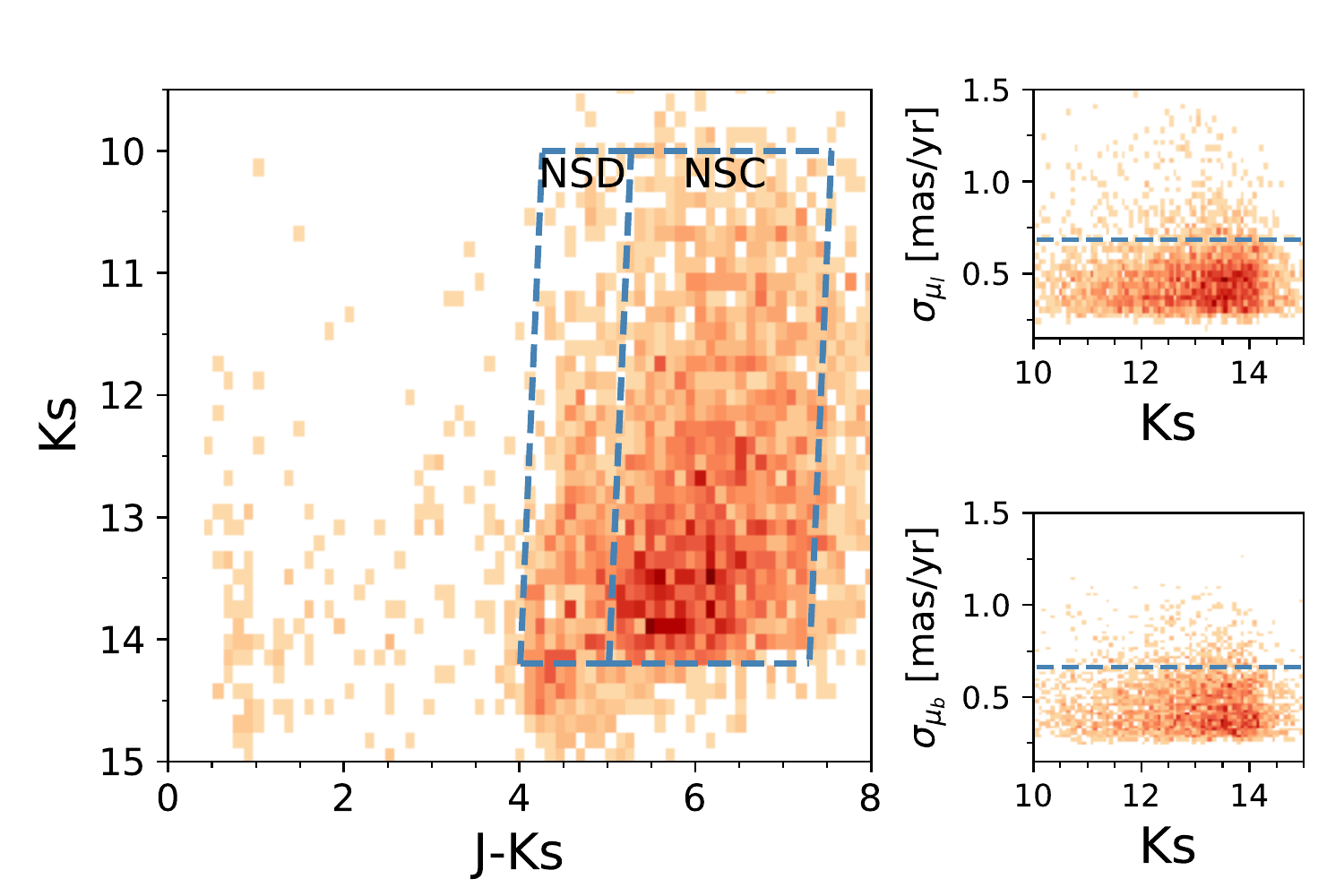}
   \caption{Colour-magnitude diagram $K_s$ versus $J-K_s$ of common stars between the photometric and the proper motion catalogues. The colour code corresponds to stellar densities using a power stretch scale. The blue dashed boxes indicate the stars belonging to the NSD and the NSC following the different extinction criterion. Right panels show the uncertainties of the proper motions $\mu_l$ (Galactic longitude coordinate) and $\mu_b$ (Galactic latitude coordinate). The blue dashed lines indicate the 90th percentile of the data (that is below 0.7\,mas/yr in both cases).}

   \label{uncer}
    \end{figure}

To analyse the proper motion distribution of each of the extinction groups, we searched for common stars between the HAWK-I photometric data and the proper motion catalogue \citep{Shahzamanian:2021wu}. Figure\,\ref{uncer} shows a CMD $K_s$ versus $J-K_s$ of the stars with available proper motions in the field. We found $\sim 7,000$ common stars and checked that more than 90\,\% of them have proper motion uncertainties below 0.7\,mas/yr for both components of the proper motions ($\mu_l$ and $\mu_b$, where $l$ and $b$ refer to Galactic longitude and latitude, respectively), as indicated in the right panels of Fig.\,\ref{uncer}.

We defined two boxes in the CMD (Fig.\,\ref{uncer}, left panel) according to the colour cut at $J-K_s\sim5$\,mag that divides the two red clump bumps with different extinction (Fig.\,\ref{CMD}). The colour cut applied to distinguish between each extinction group accounts for the predominantly old stellar population ($\gtrsim 80$\,\% of the stellar mass) that dominates the NSD and the NSC \citep{Nogueras-Lara:2019ad,Schodel:2020aa,Nogueras-Lara:2021wm}. In this way, to reduce the confusion between both extinction groups, the shape of the boxes defined in Fig.\,\ref{uncer} is in agreement with the slope of an old stellar isochrone \citep[for further details see Fig.\,4 in ][]{Nogueras-Lara:2018ab}. Moreover, the blue cut of the first box ($J-K_s\sim4$\,mag) was chosen to remove the foreground stellar population, that mainly belongs to the Galactic disc (and to some extend to the Galactic bulge/bar), and presents a significantly lower extinction than GC stars \citep{Nogueras-Lara:2021uz}. We also defined a brightness cut for each box at $K_s=14.2$\,mag, according to the detection limit of the proper motion catalogue.

\subsection{Proper motion distribution}
\label{GMM_method}

\begin{figure}
   \includegraphics[width=\linewidth]{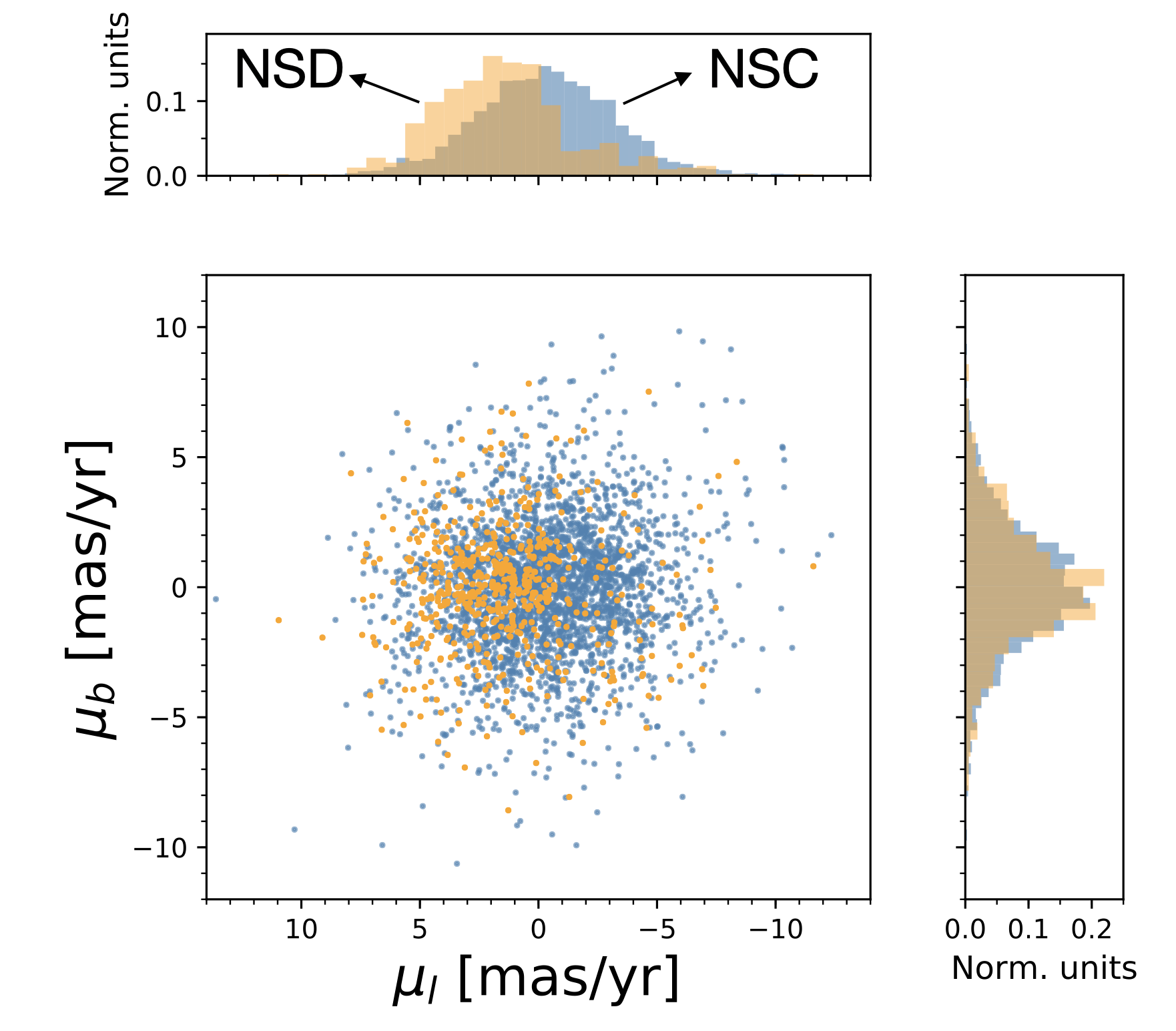}
   \caption{Proper motion distributions for the different extinction groups corresponding to the NSD and the NSC.}

   \label{proper}
    \end{figure}

Figure\,\ref{proper} shows the comparison between the proper motion distribution of each of the extinction groups (Fig.\,\ref{uncer}). We observed a clearly different distribution of the proper motions parallel to the Galactic plane ($\mu_l$) between both groups. In this way, the stars belonging to the low extinction group present a larger $\mu_l$ in comparison to those from the high extinction group. On the other hand, the distribution of the proper motion component perpendicular to the Galactic plane ($\mu_b$) appears to be similar for both groups.

\begin{figure}[h]
   \includegraphics[width=\linewidth]{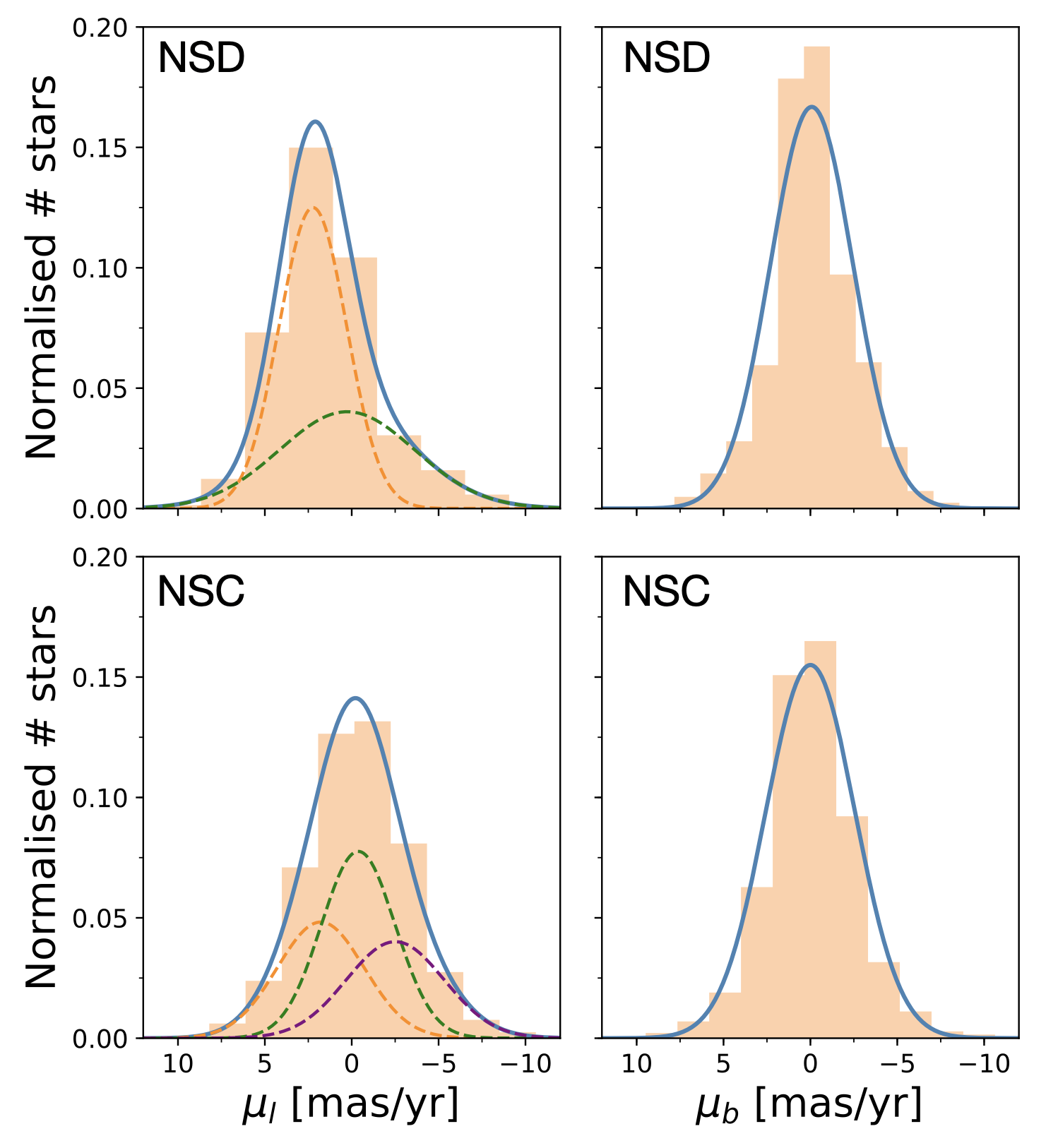}
   \caption{Upper panels: GMM modelling for the proper motion distribution ($\mu_l$ and $\mu_b$, left and right panels, respectively) of the low extinction group of stars that corresponds to the NSD. The blue lines show the total GMM model whereas the coloured dashed lines indicate each individual Gaussian contributing to the total model. Lower panels: similar to the upper panels but for the high extinction group of stars corresponding to the NSC.}

   \label{GMM}
    \end{figure}

We further analysed the proper motion distribution using the SCIKIT-LEARN Python function GaussianMixture \citep[GMM, ][]{Pedregosa:2011aa} to obtain the probability density function that originates the underlying distributions based on Gaussian models. First of all, we applied the Akaike Information criterion \citep[AIC, ][]{Akaike:1974aa} to obtain the number of Gaussian models that best describe the probability density function of the data. Following the results of \citet{Shahzamanian:2021wu} for the NSD, we tried up to three Gaussian models for the proper motion component parallel to the Galactic plane ($\mu_l$), and two models for the perpendicular component ($\mu_b$). We obtained that the $\mu_l$ distribution of the low extinction group of stars is best described by the combination of two Gaussian models, whereas a three-Gaussians model is favoured for the high extinction group. The $\mu_b$ distribution of both groups is similar and best fitted by a single Gaussian model. Figure\,\ref{GMM} shows the obtained results. To estimate the parameters and the associated uncertainties of each Gaussian model, we resorted to a Jackknife approach repeating 1,000 times the GMM modelling randomly dropping $20$\,\% of the used stars for each iteration. Table \ref{proper_motions} shows the obtained results.

\begin{table*}
\caption{Results from the GMM analysis of the proper motion distribution.}
\label{proper_motions} 
\begin{center}
\def\arraystretch{1.3}
\setlength{\tabcolsep}{3.8pt}    
    
\begin{tabular}{c|ccc|ccc}
\multicolumn{1}{c}{} & $A_l$ & $\mu_l$ & $\sigma_{\mu_l}$ & $A_b$ & $\mu_b$ & $\sigma_{\mu_b}$\tabularnewline
\multicolumn{1}{c}{} & (normalised units) & (mas/yr) & (mas/yr) & (normalised units) & (mas/yr) & (mas/yr)\tabularnewline
\hline 
\hline 
NSD & 0.61 $\pm$ 0.01 & 2.25 $\pm$ 0.10 & 1.94 $\pm$ 0.05 & - & -0.09 $\pm$ 0.05 & 2.39 $\pm$ 0.04\tabularnewline
 & 0.39 $\pm$ 0.01 & 0.23 $\pm$ 0.21 & 3.85 $\pm$ 0.16 &  &  & \tabularnewline
\hline 
NSC & 0.30 $\pm$ 0.01 & 1.80 $\pm$ 0.06 & 2.60 $\pm$ 0.05 & - & 0.01 $\pm$ 0.02 & 2.58 $\pm$ 0.02\tabularnewline
 & 0.418 $\pm$ 0.003 & -0.36 $\pm$ 0.04 & 2.12 $\pm$ 0.03 &  &  & \tabularnewline
 & 0.29 $\pm$ 0.01 & -2.50 $\pm$ 0.05 & 2.84 $\pm$ 0.08 &  &  & \tabularnewline
\end{tabular}

\end{center}
\footnotesize
\textbf{Notes.} $A_i$, $\mu_i$, and $\sigma_i$ indicate the amplitude, the mean value, and the standard deviation of each of the components of the GMM modelling, where the subindex $i$ indicates Galactic longitude ($i=l$) or latitude ($i=b$).

 \end{table*}

\subsection{Discussion}
\label{dis}

We found that the $\mu_l$ distribution of each of the extinction groups is significantly different. Therefore, their stellar populations have different kinematics, reinforcing that they correspond to different GC components that we identify as the NSD and the NSC. We explain the presence of two Gaussian models in the low extinction group because of the observation of the stellar population from the NSD that is in front of the NSC. In this way, the Gaussian component with $\mu_l = 2.25\pm0.10$\,mas/yr corresponds to the NSD stellar population moving eastwards, and agrees within the uncertainties with the value obtained by \citet{Shahzamanian:2021wu} when analysing the NSD kinematics ($\mu_l = 1.99\pm0.13$\,mas/yr). On the other hand, the secondary Gaussian ($\mu_l = 0.23\pm0.21$\,mas/yr) is probably due to stars belonging to the Galactic bulge/bar, that present a broader distribution, in agreement with previous work \citep{Clarkson:2008aa,Kunder:2012wn,Soto:2014ww,Shahzamanian:2021wu}. A third Gaussian component corresponding to the stellar population moving westwards \citep[e.g.][]{Shahzamanian:2021wu} is not visible due to the presence of the NSC, due to its high stellar density and also the increase of interstellar reddening towards the farther edge of the NSD.

We interpret the three Gaussian components detected in $\mu_l$ distribution of the high extinction stellar group, as the result of the rotation of the NSC that produces two main peaks due to stars moving eastwards ($\mu_l = 1.80\pm0.06$\,mas/yr), and westwards ($\mu_l = -2.50\pm0.05$\,mas/yr). The third peak centred on $\mu_l = -0.36\pm0.04$\,mas/yr is probably due to the differential rotation of the NSC. In this way, stars rotating slower than the eastwards and westwards population producing the previously described peaks, are likely originating the central one. This is compatible with the presence of a radial dependence of the rotation velocity, that was measured to be smaller ($\sim 0.5$\,mas/yr) for the innermost parsec of the NSC  \citep{Trippe:2008it,Schodel:2009zr}. Moreover, some contamination from the NSD and the Galactic bulge/bar might also contribute to the observed distribution. However, due to the higher density of stars present in the NSC in comparison with the NSD, in combination with the significantly larger extinction due to the given colour cut, we expect that the contamination from the Galactic bulge/bar will be less important for this stellar population than for the previous case of the NSD. 

The fact that the third peak is not centred on zero is probably because the proper motions in the used catalogue were computed as relative ones assuming that the mean proper motion of all the stars in a given field is zero \citep{Shahzamanian:2021wu}. In this way, due to the presence of the NSC in the studied region, it is not possible to observe stars from the far side of the NSD, and thus the reference frame for $\mu_l$ is dominated by a majority of stars moving eastwards. Hence, the positive component of $\mu_l$ is underestimated, whereas the negative one is overestimated. This scenario is consistent with the larger absolute value of $\mu_l$ obtained for the stars moving westwards in comparison with the ones moving eastwards ($\mu_l = -2.50\pm0.05$\,mas/yr, and $1.80\pm0.06$\,mas/yr, respectively). Actually, assuming that the $\mu_l$ component of the proper motions is shifted $\mu_l = +0.36\pm0.04$\,mas/yr, we obtain that the components due to the rotation of the NSC are equal in absolute value within the uncertainties, and that the third Gaussian peak is centred on zero.

Analysing the $\mu_l$ distribution of the NSC for different colour cuts, we checked that the stellar population moving westwards is more extinguished. This is compatible with the lower number of stars belonging to this Gaussian component obtained in the GMM modelling (Table\,\ref{proper_motions}), which is probably due to the extinction by dust within the NSC \citep{Chatzopoulos:2015uq}. Therefore, we obtained that the direction of rotation is the same for both the NSC and the NSD, in agreement with previous work \citep[e.g.][]{Feldmeier:2014kx,2015ApJ...812L..21S,Shahzamanian:2021wu}. Moreover, the values obtained for the eastwards moving component of the NSD and the NSC indicate a faster rotation of the NSD in comparison to the NSC, as it was expected \citep[e.g. Fig.\,18 in ][]{Sormani:2022wv}.

On the other hand, both extinction groups show similar distributions of $\mu_b$ (except for a somewhat larger dispersion for stars from the high extinction group), that can be best reproduced by a single Gaussian model approximately centred on zero. Although the presence of some contamination from the Galactic bulge/bar in the low extinction group (NSD) might have originated a two Gaussians distribution \citep{Shahzamanian:2021wu}, this was probably not observed due to the low number of stars from the Galactic bulge/bar in the sample. However, we measured a broadening of the detected single Gaussian component ($\sigma_{\mu b} = 2.39\pm0.05$\,mas/yr) in comparison with the expected one for the NSD according to previous work \citep[$\sigma_{\mu b} = 1.499\pm0.0002$\,mas/yr, ][]{Shahzamanian:2021wu}, that is probably due to the influence of the expected Galactic bulge/bar stars. This is in agreement with the results of \citet{Sormani:2022wv}, that estimated the contamination of stars from the Galactic bulge/bar for different fields distributed across the NSD. They obtained that the contribution from Galactic bulge/bar stars is less important for their fields closest to the NSC ($\lesssim20$\,\% of the stars with $H-K_s>1.3$\,mag belong to the Galactic bulge/bar stars, see Table\,2 in \citealt{Sormani:2022wv}). Although the regions analysed in \citet{Sormani:2022wv} avoid the NSC, that is precisely the target field of this work, we expect that the relative contribution of Galactic bulge/bar will be lower than in their innermost fields, due to the presence of a significantly higher number of stars belonging to the NSC. On the other hand, in this work we use the near infrared bands $J$ and $K_s$, that allow us to remove the foreground stellar population (from the Galactic disc and also from the bulge/bar), in a more efficient way than the $H$ and $K_s$ bands used by \citet{Sormani:2022wv}, due to the longer wavelength base line.

\section{Radial velocities}

We analysed the distribution of radial velocities of spectroscopically characterised stars from \citet{Feldmeier-Krause:2017kq,Feldmeier-Krause:2020uv}. We searched for common stars between our photometric catalogue and the stars with known metallicities ($\sim 1,000$ common stars), and classified them following the boxes defined in Fig.\,\ref{CMD}. We obtained that $\lesssim 10$\,\% of the stars with known metallicities belong to the low extinction group. We used a photometric criterion instead of a proper motion analysis due to the low number of common detections between the proper motion catalogue and the stars with known metallicities (see the gaps of the proper motion catalogue in Fig.\,\ref{GNS}). 

\begin{figure}[h]
   \includegraphics[width=\linewidth]{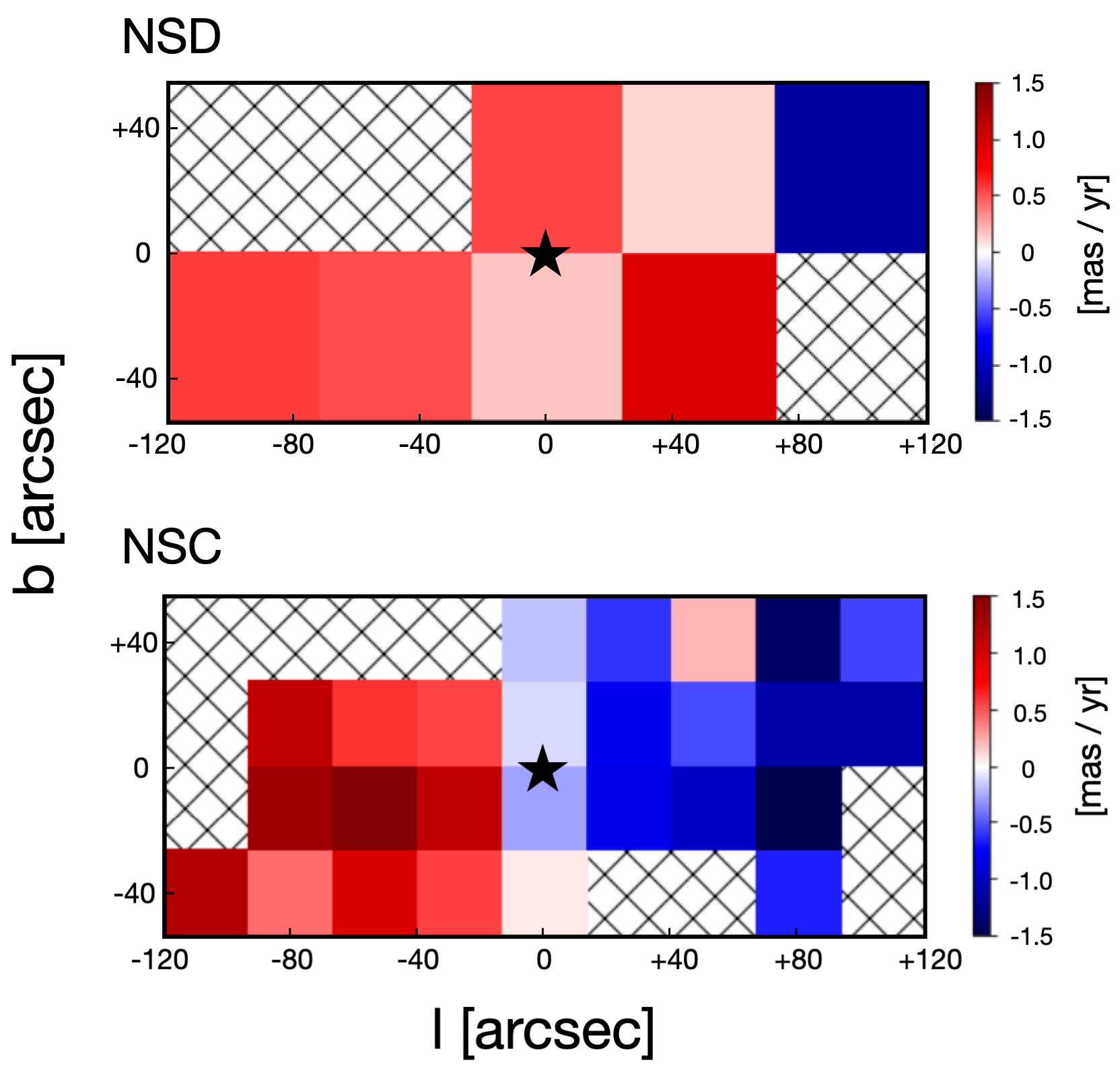}
   \caption{Radial velocity maps obtained for the NSD (upper panel) and the NSC (lower panel). The x- and y-axes indicate the distance in arcseconds from Sagittarius A* (indicated by a black star). Cross-shaped pixels indicate that there are not enough stars to compute a radial velocity value. The pixel size is larger in the upper panel due to the low number of NSD stars in the region ($\sim60$ stars).}

   \label{vr}
    \end{figure}

We built a radial velocity map for the stars belonging to each of the extinction groups. We defined a pixel size of $\sim 50''$, and $\sim 25''$ for the radial velocity maps corresponding to the NSD and the NSC, given the significantly lower number of stars detected for the NSD (around 10\,\% of the total number of stars). We computed the radial velocity for a given pixel calculating the mean value of the radial velocities from the stars located within the pixel imposing a 3-sigma clipping algorithm to remove outliers. We only computed a value for a given pixel if at least 4 stars were present within it. Figure\,\ref{vr} shows the obtained maps. The radial velocity distribution appears to be different between the NSD and the NSC. In this way, the NSD map is dominated by positive values. This might be due to the NSD geometry that is not well known yet, the low number of stars to produce this map ($\sim 60$\, stars), and also the possible contamination from the Galactic bulge/bar that can bias the results. On the other hand, the NSC map clearly shows a rotation pattern that is also compatible with the differential rotation suggested in Sect.\,\ref{dis} to explain the three Gaussian components observed for the $\mu_l$ distribution.

\section{Metallicity}

\begin{figure}[h]
   \includegraphics[width=\linewidth]{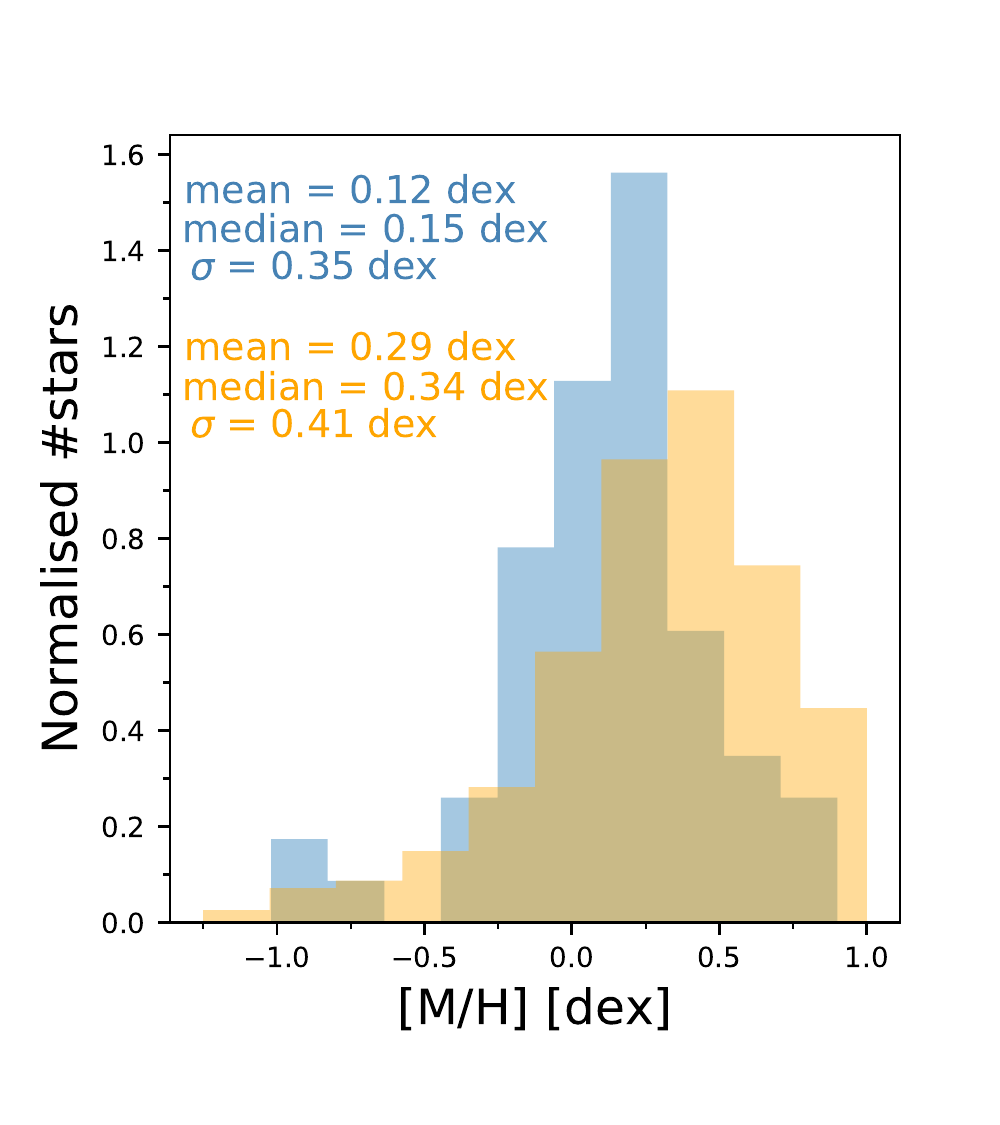}
   \caption{Metallicity distribution of the stars belonging to the NSD (in blue), and the NSC (in orange). The mean, the median, and the standard deviation of each distribution are shown in the Figure. The systematic uncertainty of the metallicity measurements is $\sim0.25$\,dex according to \citet{Feldmeier-Krause:2017kq,Feldmeier-Krause:2020uv}.}

   \label{Met}
    \end{figure}

We studied the metallicity distribution of the stars in the target region, using the previously computed list of common stars between our photometric catalogue and stars with known metallicities \citep{Feldmeier-Krause:2017kq,Feldmeier-Krause:2020uv}. Figure\,\ref{Met} represents the metallicity distribution of the stars corresponding to each of the extinction groups. Our results show different metallicity distributions for each extinction group, being the high extinction group more metal rich than the low extinction one. This indicates the presence of two components with different metallicity and agrees with the results obtained for the NSD and the NSC \citep[e.g.][]{Feldmeier-Krause:2017kq,Feldmeier-Krause:2020uv,Fritz:2020aa,Schultheis:2021wf,Feldmeier-Krause:2022vm}. The NSC metallicity obtained after removing the contribution from the NSD is somewhat higher than the obtained without considering the presence of the NSD \citep[as done in previous work, e.g.][]{Feldmeier-Krause:2017kq}. This is because NSD stars are less metal rich in average and bias the NSC sample when they are not removed.

\begin{figure}
   \includegraphics[width=\linewidth]{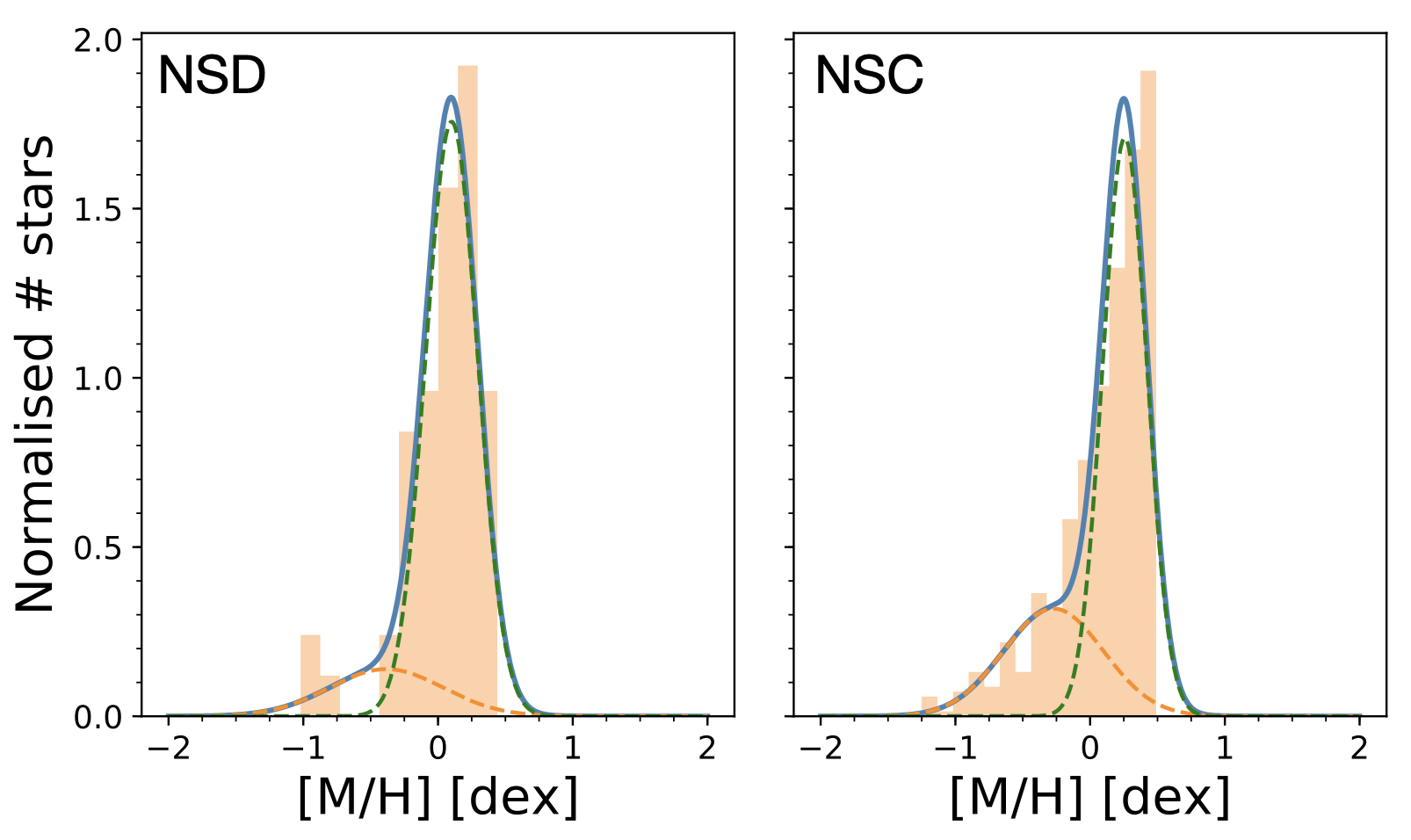}
   \caption{Best GMM model obtained for the stars with known metallicities belonging to each of the defined extinction groups corresponding to the NSD (left panel), and the NSC (right panel). The blue lines show the total GMM model whereas the coloured dashed lines indicate each individual Gaussian contributing to the total model.}

   \label{Met_GMM_fig}
    \end{figure}

We further study the metallicity distribution of each component restricting the analysis to stars with metallicity [M/H] < 0.5\,dex. In this way, we removed stars whose metallicity might have been overestimated due to the calibration used in \citealt{Feldmeier-Krause:2017kq,Feldmeier-Krause:2020uv} (see Sect.\,\ref{metal}). We applied the GMM method previously explained (Sect.\,\ref{GMM_method}) to each of the extinction groups. We used the AIC to distinguish between models considering one and two Gaussians. We obtained that the metallicity distribution of both extinction groups is best represented by a two-Gaussians model  (Fig.\,\ref{Met_GMM_fig}). To estimate the mean value and uncertainty of each component, we generated 1,000 Monte Carlo samples for the metallicity distribution, assuming Gaussian uncertainties for the metallicity of each star and randomly varying its value considering a Gaussian distribution. Table\,\ref{Met_GMM} shows the obtained results, where the mean and the standard deviation were determined averaging over the results of the 1,000 Monte Carlo samples, applying a 3-sigma outlier-resistant criterion to remove outliers.

For both extinction groups the metallicity distributions present a broad component centred around [M/H] $\sim$ -0.20\,dex, and a more metal rich narrower one, whose metallicity is larger for the NSC stellar population. Our results agree with the recent work by \citet{Schultheis:2021wf}, where two stellar components with different metallicities were found for both, the NSC and the NSD, probably indicating a different formation for each of these two components. Actually, part of the data used in \citet{Schultheis:2021wf} to analyse the NSC corresponds to the same data set that we used in our analysis \citep{Feldmeier-Krause:2020uv}. Nevertheless, \citet{Schultheis:2021wf} assumed all the stars in the sample to be part of the NSC. Here, we checked that $\lesssim 10$\,\% of the stars that we used belong to the NSD and have a lower mean metallicity in comparison to the stars from the NSC. Therefore, considering them as NSC stars might slightly bias the results towards lower metallicity values for the NSC. On the other hand, we are able to disentangle the stellar populations from the NSD and the NSC using the same data set and line of sight for both components, reinforcing the results obtained in \citet{Schultheis:2021wf} when comparing different data sets and lines of sight.

Previous high-resolution spectroscopic studies have been carried out for some stars in the NSC. In particular, \citet{Ryde:2016wr} found a metal-poor ($[Fe/H]\sim-1.0$\,dex), red giant star at a projected distance of 1.5\,pc from Sagittarius\,A*, and determined that it probably belongs to one of the nuclear components (NSD or NSC), confirming the presence of metal poor stars in the GC. On the other hand, \citet{Do:2018aa} targeted two stars found to be very metal rich in previous spectral-template fitting work \citep{Do:2015ve}. They confirmed that at least one of them has an unusually high metallicity $[M/H]>0.6$\,dex, proving the presence of these kind of stars in the GC. Moreover,  \cite{Rich:2017rm} determined the metallicity of 17 stars in the NSC to span between $-0.5 < [Fe/H] < +0.5$\,dex, being these values also compatible with the range obtained in this work. Finally, the recent work by \cite{Thorsbro:2020uq} found that the most metal rich stars in their NSC sample reach $[Fe/H] = +0.5$\,dex. These values end up in higher metallicities when considering the overall metallicity, $[M/H]$, that are in agreement with the most metal rich stars that we use from \citet{Feldmeier-Krause:2017kq,Feldmeier-Krause:2020uv}. We would like to stress that the use of $JK_s$ photometry and proper motions can be very helpful to reanalyse previous work where some of the stars that were considered to belong to the NSC might be part of the NSD.

\begin{table}
\caption{Results from the GMM analysis of the proper motion distribution.}
\label{Met_GMM} 
\begin{center}
\def\arraystretch{1.3}
\setlength{\tabcolsep}{3.8pt}    
   
    \begin{tabular}{cccc}
\hline 
Component & $A_{[M/H]}$ & $[M/H]$ & $\sigma_{[M/H]}$ \tabularnewline
\hline 
\hline 
NSD & $0.38$ $\pm$ $0.13$ & $-0.22$ $\pm$ $0.16$ & $0.47$ $\pm$ $0.15$\tabularnewline
 & $0.62$ $\pm$ $0.13$ & $0.17$ $\pm$ $0.08$ & $0.30$ $\pm$ $0.05$\tabularnewline
\hline 
NSC & $0.40$ $\pm$ $0.02$ & $-0.17$ $\pm$ $0.04$ & $0.49$ $\pm$ $0.03$\tabularnewline
 & $0.60$ $\pm$ $0.02$ & $0.28$ $\pm$ $0.02$ & $0.31$ $\pm$ $0.01$\tabularnewline
\hline 
 &  &  & \tabularnewline
\end{tabular}
  \end{center}
\footnotesize
\textbf{Notes.} $A_{[M/H]}$, $[M/H]$, and $\sigma_{[M/H]}$ indicate the amplitude, the mean value, and the standard deviation of each of the components of the GMM modelling.

 \end{table}

\section{Conclusion}  
  
In this paper we analysed the kinematics and metallicity of the stars belonging to two extinction groups identified along the line of sight towards the NSC \citep{Nogueras-Lara:2018aa,Nogueras-Lara:2021wm}. We detected two kinematically distinct components associated to the each of the extinction groups, that we confirmed as the NSD and the NSC.  In this way, our results show the potential of proper motions to disentangle the stellar population belonging to each of the GC structures. We analysed their proper motion components parallel ($\mu_l$) and perpendicular ($\mu_b$) to the Galactic plane, and found that the $\mu_l$ distributions of the NSD and the NSC are best fitted by two and three Gaussian models, respectively. In this way, we explained the NSD proper motion distribution as the combination of the stellar population from the closest edge of the NSD (that is rotating eastwards), and some contamination from Galactic bulge/bar stars that cannot be easily removed from the sample due to their large extinction. We concluded that the presence of the NSC impedes the detection of stars from the far side of the NSD, that are rotating westwards \citep{Shahzamanian:2021wu}. The $\mu_l$ distribution of the NSC shows three Gaussian components that we explain as a consequence of the rotation of the NSC. These components correspond with stars moving eastwards (positive $\mu_l$) and westwards (negative $\mu_l$), and stars moving with relatively slower velocities from the innermost regions of the NSC. We obtained relatively lower values for the rotation of the NSC in comparison with the NSD, as it was expected \citep[e.g.][]{Sormani:2022wv}. The $\mu_b$ distributions of both, the NSD and the NSC, seem to be similar except for a somewhat higher dispersion for the NSC values.
 
We created radial velocity maps using spectroscopically characterised stars from each of the extinction groups and found that the velocity distributions seem to be different between the NSD and the NSC. On the other hand, we observed a velocity pattern for the NSC compatible with differential rotation.

We also analysed stars with known metallicities from each of the extinction groups. We found that they follow different distributions, being the group corresponding to the NSD less metal rich, in agreement with previous studies \citep[e.g.][]{Feldmeier-Krause:2017kq,Feldmeier-Krause:2020uv,Fritz:2020aa,Schultheis:2021wf,Feldmeier-Krause:2022vm}. We obtained that both components can be best described by a two-Gaussians model with a less metal-rich wider component, and a predominant metal rich narrower one, that might have a different origin in agreement with \citet{Schultheis:2021wf}. Moreover, we measured a mean value of the NSC metal rich stellar population of $[M/H] \sim 0.3$\,dex, that arguably makes the NSC the most metal rich region of the Galaxy.

Our results confirm that the NSC and the NSD can be distinguished along the line of sight towards the NSC via their different extinction and agree with previous work on the NSD and the NSC suggesting different formation scenarios and stellar populations \citep[e.g.][]{Nogueras-Lara:2019ad,Schodel:2020aa,Schultheis:2021wf,Feldmeier-Krause:2022vm,Nogueras-Lara:2021wm}.

  \begin{acknowledgements}
F. N.-L. gratefully acknowledges the sponsorship provided by the Federal Ministry for Education and Research of Germany through the Alexander von Humboldt Foundation. This work is based on observations made with ESO Telescopes at the La Silla Paranal Observatory under program IDs 60.A-9450(A), 091.B-0418, 093.B-0368,  and 195.B-0283. F. N.-L. thanks Rainer Sch\"odel, Nadine Neumayer, and Mattia Sormani for very useful discussion.

\end{acknowledgements}

\bibliography{BibGC.bib}
\end{document}